\documentclass[useAMS,usenatbib]{mn2e}
\usepackage[utf8]{inputenc}
\usepackage{graphicx}
\usepackage{amsmath}
\usepackage{pifont}
\usepackage{multirow}
\usepackage{verbatim}

\title[Red-shifted diffuse bands]{Red\textendash shifted Diffuse Interstellar
Bands in Orion OB1 association
}

\author[Kre{\l}owski et al.]
{J. Kre{\l}owski$^{1}$,  G.A. Galazutdinov$^{2,3}$, G.
Mulas$^{4}$, M. Maszewska$^{1}$,
\and C. Cecchi-Pestellini$^{5}$\\
$^{1}$Center for Astronomy, Nicholas Copernicus University,
Gagarina 11, Pl-87-100 Toru{\'n}, Poland\\
$^{2}$Instituto de Astronomia,
       Universidad Catolica del Norte,
       Av. Angamos 0610, Antofagasta
       Chile\\
$^{3}$Pulkovo Observatory, Pulkovskoe Shosse 65, Saint-Petersburg 196140, Russia\\
$^{4}$INAF - Osservatorio Astronomico di Cagliari, via della
scienza 5 - 09047 Selargius, Italy \\
$^{5}$INAF \textendash Osservatorio Astronomico di Palermo, piazza Parlamento 1
\textendash 90134 Palermo, Italy}

\begin{document}

\date{Accepted . Received ; in original form }

\pagerange{\pageref{firstpage}--\pageref{lastpage}} \pubyear{2014}

\maketitle

\label{firstpage}

\begin{abstract}
The wavelength displacement of the Diffuse Interstellar Bands
at 4502, 5705, 5780, 6284, and 7224~\AA\ with respect to the well known,
narrow atomic/molecular interstellar lines (of Ca{\sc ii} and
Na{\sc i}) have been measured in the spectra of the 2 Orion
Trapezium stars HD~37022 and HD~37020, using the HARPS\textendash N
spectrograph, fed with the 3.5~m Telescopio Nazionale Galileo, and
the BOES spectrograph, fed with the 1.8m Korean telescope.
The red\textendash shift is $\sim$25~km/s for all these DIBs.
We discuss the various possible origins of this very peculiar
wavelength shift in the light of the particular physical conditions
in the Orion Trapezium. The above mentioned shift is seemingly
absent in the DIBs at 6196 and 6993~\AA.
\end{abstract}


\section{Introduction}
Absorption spectra of interstellar clouds contain well\textendash
known molecular bands of simple polar radicals, CH, CN, CH$^+$
which have been discovered and identified long ago \citep{kellar41}.
For a long period they were believed to be the only
possible interstellar molecules. In 1970-ties small homonuclear
molecules (H$_2$, C$_2$) were also found. At the same time,
rotational emission features revealed the presence (specifically
in the dense ISM, mostly in star\textendash forming regions) of
many complex molecules with a dipole momentum; the full list of
these polar species (mostly carbon\textendash bearing) currently
counts more than 190 entries\footnote{see e.~g.
http://astrochymist.org/astrochymist\_ism.html},
clearly demonstrating that a rich chemistry, mostly
carbon\textendash based, takes place in interstellar clouds.
Some of these species have recently
been observed in the middle and far infrared by the space
telescope Herschel. The largest currently known interstellar
molecules are the fullerenes C$_{60}$ and C$_{70}$
\citep{cami10,berne2013}. Yet unknown (unidentified) molecular
species are believed to be the carriers of diffuse interstellar
bands (DIBs): these make up the longest standing unsolved problem
in astronomical spectroscopy. The first two DIBs were discovered
in 1921 by \citet{heger22}. Until 1970 the list of DIBs was very
short \textemdash\ only 9 entries. The application of solid state
detectors and more powerful telescopes to DIB observations led to
the progressive discovery of many additional weak features, almost
all of them in the range from the near\textendash UV to the
near\textendash IR. The current lists of known DIBs contain over
400 entries \citep{hobbs09, GKMEF00}, the majority
of them is very shallow.
Some of the weaker features are not unambiguously detected and
classified as DIBs in all surveys, e.g. \citet{bondar12}.
Even more importantly, fine structures \textemdash\ reminiscent in
some cases of the unresolved rotational envelopes of bands of polyatomic
molecules \textemdash\ have been resolved in some DIBs \citep{kerr98}.
Nearly all conceivable forms of matter, from the
hydrogen anion to color centres in dust grains, have
already been proposed as DIB carriers. All of them have been rejected,
with very few still tentative exceptions. Recently \citet{krel10}
found a very weak DIB which coincides with an electronic, gas\textendash phase
band of HC$_4$H$^+$, supporting the molecular DIB origin.
Another coincidence between a lifetime\textendash broadened absorption
spectrum recorded through a hydrocarbon plasma and a strong DIB at 545~nm
was reported by \citet{linn10}, but an unambiguous identification of the
carrier has not yet been possible. l\textendash C$_3$H$_2$ was
initially proposed by \citet{maier2011} but this identification
would imply this species to be surprisingly abundant in the ISM;
moreover, another broad DIB, initially interpreted as a
second l\textendash C$_3$H$_2$ feature (from the same level) near
4883\AA, clearly has a different origin \citep{krel11}.
Two strong DIBs in the near\textendash IR were tentatively
attributed to C$_{60}^+$ by \citet{FE94} on the
basis of laboratory spectra measured in solid rare\textendash gas
matrices, but this was disputed by \citet{JMPB97} and
\citet{GKMEF00}.
Despite the recent detection of the vibrational spectrum of
C$_{60}^+$ \citep{berne2013}, its identification or
rejection as DIB carrier still awaits gas\textendash phase
laboratory spectra, as yet unavailable. Similarly,
\citet{IGMGH08,IGGHM12} recently claimed the
identification of electronic transitions of gas\textendash phase
singly charged naphthalene, the simplest Polycyclic Aromatic Hydrocarbon
(PAH) in the spectra of Cernis~52 and, more recently, of HD~125241.
In Cernis~52 a well\textendash defined band is firmly detected, which
is consistent with the strongest band of C$_{10}$H$_8^+$. Other, weaker
bands are compatible with observations, but their presence is not
as firmly established, mainly due to stellar line contamination.
In the case of HD~125241, the authors may have been misled by the fact
that the spectrum of this star is populated with many stellar emission
lines, possibly of circumstellar origin. One of them, centered at
6701.2~\AA\ and due to Si{\sc iv}, is very broad. Setting the
continuum over the red wing of this emission can easily have created
a broad depression as an artifact, where the authors identify the
strongest naphthalene band. This was never confirmed in any other
sources, nor any other features of this molecule have been
firmly identified yet on the same source, being at best weak features
superimposed on stronger stellar features contaminating them.
\citet{IGMRGHGHL10} also claimed the identification of a band of
the anthracene cation (C$_{14}$H$_{10}^+$) in the spectrum of Cernis~52.
However, this one band is obtained after subtraction of a stronger,
very close, overlapping stellar feature, and no other bands of
C$_{14}$H$_{10}^+$ were detected. It was furthermore rebutted by
\citet{Gala11}, so that this claimed identification must be regarded as
tentative. No other specific PAH features were found, despite
several dedicated efforts \citep[see e.~g.][]{salama2011, gredel2011}.

\begin{table*}
\caption{Target list with basic stellar data
\citep[from the SIMBAD database,][]{simbad2000}.}
\label{targetlist}
\begin{tabular}{|c|c|c|c|c|c|c|c|c|} \hline
HD number & RA & Dec & l & b & Sp. type & V (mag) & E$_\mathrm{B-V}$ & {v}$_\mathrm{rad}$ (km/s)\\ \hline
22928 & 03 42 55.504 & +47 47 15.17 & 150.2834 & -05.7684 & B5III C & 3.01 & 0.00 & 4.0 km/s\\ \hline
24398 & 03 54 07.922 & +31 53 01.08 & 162.2891 & -16.6904 & B1Ib & 2.85 & 0.31 & 20.6 km/s\\ \hline
37020 & 05 35 15.829 & -05 23 14.36 & 209.0072 & -19.3853 & B0.5V & 6.73 & 0.26 & 28.3 km/s\\ \hline
37022 & 05 35 16.464 & -05 23 22.85 & 209.0107 & -19.3841 & O7V & 5.13 & 0.31 & 23.6 km/s\\ \hline
\end{tabular}
\end{table*}

After more than 90 years of DIB research, despite having learned much
about diffuse and translucent clouds, the identification of DIBs is
thus still missing.
We emphasize that a vast majority of reddened stars is observed
through several translucent clouds. As a result we observe DIBs in
ill\textendash defined averages \textemdash\ physical parameters
and spectra of individual clouds may be widely different \citep{KW88}.
The fact that DIBs are most frequently observed through
a number of possibly very different clouds is likely to be one of the reasons
for their apparent uniformity, washing out individual differences.
This conversely makes all ``peculiarities'' very
attractive, as the latter are most likely spectra of individual
clouds, where the physical conditions may be more homogeneous than
in a random sample of clouds and thus a more sensible, unambiguous physical
interpretation of the spectra is possible. The examples of extreme
``peculiarities'' shown by \citet{KW88} \textemdash\ namely $\sigma$Sco and
$\zeta$Oph \textemdash\ became the archetypes of translucent interstellar
clouds of evidently different spectra.

In this sense, objects belonging to Ori OB1 association, a region of
active star formation, were discovered to be among the most interesting
ones. The stars forming the famous Orion Trapezium are quite substantially
reddened, but the clouds obscuring these objects are very
peculiar. Their extinction curves are very unusual \citep{FM07}
and the features carried by the simple radicals
typically fall below the level of detection. For what concerns
DIBs, only two pretty broad ones, at 5780 and 6284~\AA, are strong
with respect to extinction; the ones at 6196, 6205, 6010, and
5705~\AA\ are detectable but weak \citep{KG99}.
\citet{KG99} in the same paper also
found rather surprising red\textendash shifts of some DIBs in relation
to the interstellar sodium doublet, the only evident atomic line
clearly seen in the then available spectra of the Ori OB1 stars.
This result was confirmed later by \citet{WWBK01}, but only
for HD~37061. Since the stability of the rest wavelengths of DIBs is
one of their most well\textendash established observational
properties, this result is indeed extremely peculiar, and we
therefore set out to further investigate it.

The present paper aims to confirm the above results and to relate
the peculiar behaviour of DIBs to that of simple molecular
species. The point is to use high quality, high resolution spectra
from more than one instrument. The Ori OB1 association
is the nearest region of active star formation and thus it is very
interesting what kind of physical conditions may be caused by the
proximity of clouds and young, hot stars and how these conditions
may influence the formation of DIB carriers, DIB spectral profiles
and wavelength shifts.

\section{Observations}

We have used the HARPS\textendash N echelle
spectrograph\footnote{http://www.tng.iac.es/instruments/harps/} 
\citep{Cosentino2012},
fed with the 3.5~m Telescopio Nazionale Galileo. It offers a
resolution of R=115,000 in the spectra, divided into 70 orders.
This resolution allows to resolve Doppler components likely present in
interstellar features, while the telescope size allows to achieve a
high S/N ratio. Moreover, having been designed and built specifically
for the main purpose of detecting exoplanets measuring radial
velocity variations via accurate determinations of Doppler shifts,
HARPS\textendash N has a very stable and reliable wavelength calibration.

In addition, to be able to rule out instrumental artifacts, we
also used data of the same targets collected by the
BOES echelle
spectrograph\footnote{http://www.boao.re.kr/BOES/BOESppt3.files/frame.htm}
\citep{Kim2007} attached to the 1.8m telescope of the Bohyunsan 
Observatory in Korea.
This instrument can operate in three observational modes providing
resolving powers of 30,000, 45,000 and 90,000 respectively.
Our spectra were taken using the highest one, enabling us to
observe relatively weak, narrow features. In all
observational modes the spectrograph records the whole spectral
range from $\sim$3500 to $\sim$10,000 \AA\, divided into 75\textendash
76 spectral orders.

All of the data were reduced using IRAF \citep{tody1986} and the DECH
data reduction suite, authored and made
available\footnote{http://www.gazinur.com/DECH-software.html}
by one of the authors (G.~Galazutdinov). Data reduction involved
the standard procedure of order location, bias and background
subtraction, flatfielding, extraction of 1\textendash D spectra
order by order, and wavelength calibration
through comparison with a reference spectrum of a Th\textendash Ar
lamp. The accuracy of the wavelength calibration and the absence
of instrumental effects, which can displace the final wavelength
solution, were verified by the cross\textendash correlation of the
strong telluric bands present in all studied targets with a reference
synthetic telluric spectrum.

The stars we observed are HD~37022 and HD~37020, both in the Orion Trapezium,
HD~24398 as a reference ``normal DIBs'' star, and the bright unreddened
star HD~22928, used as a telluric line divisor. Their basic data are
given in Table~\ref{targetlist}.

\section{Results}

HARPS\textendash N allows very accurate and reliable determinations of
wavelengths and this is why we used it in this project. The range covered
by the spectrograph (from 3855 to 6910~\AA) includes two prominent 
doublets of Ca{\sc ii} and Na{\sc i} \textemdash usually very strong. The 
Orion Trapezium stars are also known to exhibit strong nebular emission lines
\citep[see e.~g.][]{baldwin2000},
apparently originating in close vicinities of the observed stars.
This spectrograph is thus very well\textendash suited to check whether the
red\textendash shift of some DIBs reported by \citet{KG99} can be confirmed.

\begin{figure}
\includegraphics[angle=0, width=\hsize]{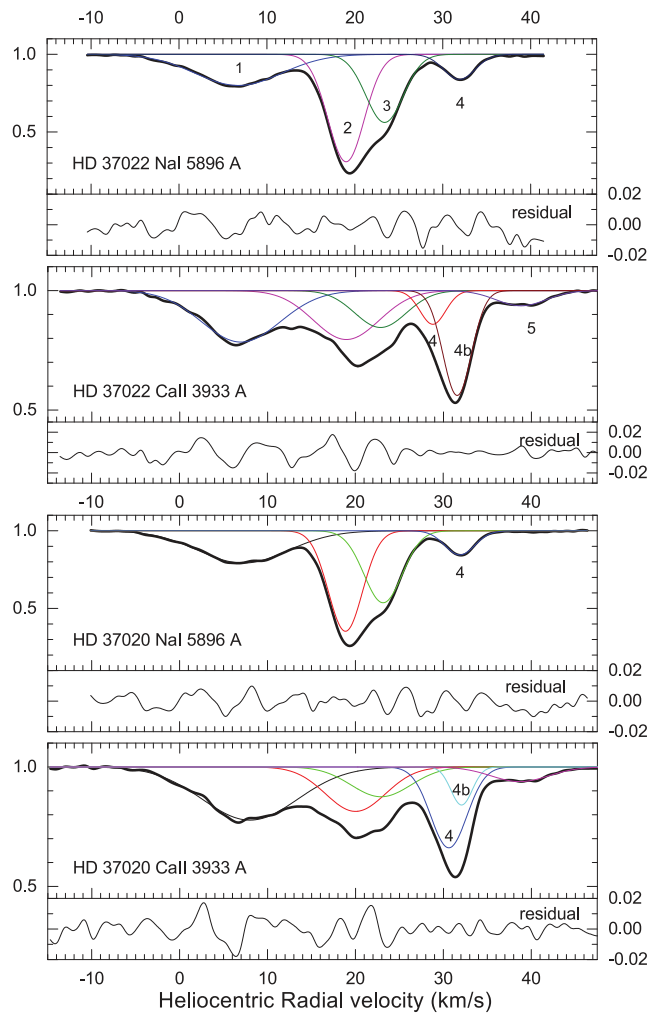}
\caption{Doppler structure of the atomic lines of Ca{\sc ii} and
Na{\sc i} lines, in radial velocity scale, for HD~37022 (top panel) and
HD~37020 (bottom panel). In each spectrum the thick black line is the 
normalised observed flux, thin coloured lines are individual fitted 
components, the residual (difference between observation and fitted model) 
is shown below.}\label{atom_doppler}
\end{figure}
Figure~\ref{atom_doppler} shows a zoom\textendash in of the two atomic lines
of Na{\sc I} at 5896~\AA\ and Ca{\sc ii} at 3933~\AA\, in heliocentric radial
velocity scale, for the Orion trapezium stars HD~37020 and HD~37022. These
stars are separated by less than 13'', and indeed the lines of sight towards
both of them appear to traverse very nearly the same foreground ISM, showing
almost identical components with the same radial velocities and equivalent
widths.
Not unusually, while both Na{\sc i} and Ca{\sc ii} show matching components at
almost the same radial velocities, intensity ratios are rather different,
their respective absolute maxima are in different radial velocity components.

Diffuse band Doppler components usually have a better
correlation with neutral rather than with ionized species \citep{GMPK04},
hence in the following we considered the strongest
Na{\sc i} component as the reference velocity for DIBs.

\begin{table}
\caption{Heliocentric radial velocity components measured in the Na{\sc I} line
at 5896~\AA\ and Ca{\sc II} line at 3933~\AA\ in the two Orion Trapezium stars
HD~37022 and HD~37020. Radial velocities are given in km/s, and their error
is less than 0.5~km/s. Equivalent widths are in m\AA.} 
\label{atomic_radvel_table}
\begin{center}
\begin{tabular}{lccc}
\hline
\hline
                          &            & HD~37022            & HD~37020 \\
&                         & \multicolumn{2}{c}{rad.~vel. (eq.~width)}  \\
\hline
\multirow{2}{*}{Comp.~1}  & Na{\sc I}  & 6.8 (48.7$\pm$1.3)  & 7.2 (52.6$\pm$1.8) \\
                          & Ca{\sc II} & 7.0 (31.7$\pm$0.9)  & 8.0 (39.74$\pm$1.3) \\
\hline
\multirow{2}{*}{Comp.~2}  & Na{\sc I}  & 19.0 (70.4$\pm$0.6) & 18.7 (61.8$\pm$0.7) \\
                          & Ca{\sc II} & 19.0 (25.2$\pm$0.8) & 20.0 (20.8$\pm$1.1) \\
\hline
\multirow{2}{*}{Comp.~3}  & Na{\sc I}  & 23.4 (46.0$\pm$0.6) & 23.1 (49.6$\pm$0.8) \\
                          & Ca{\sc II} & 22.8 (14.9$\pm$0.7) & 22.9 (13.9$\pm$0.8) \\
\hline
\multirow{2}{*}{Comp.~4}  & Na{\sc I}  & 31.9 (14.2$\pm$0.5) & 31.9(13.0$\pm$0.6) \\
                          & Ca{\sc II} & 28.8 (6.7$\pm$0.3)  & 30.6 (22.8$\pm$0.5) \\
Comp.~4b                  & Ca{\sc II} & 31.6 (23.4$\pm$0.4) & 32.1 (6.4$\pm$0.3) \\
\hline
\multirow{2}{*}{Comp.~5}  & Na{\sc I}                        & \textemdash & \textemdash \\
                          & Ca{\sc II} & 38.8 (5.9$\pm$0.6)  & 38.7 (6.6$\pm$0.7) \\
\hline \hline
\end{tabular}
\end{center}
\end{table}

Table~\ref{atomic_radvel_table} lists the measured components, both in Na{\sc i}
and Ca{\sc ii} lines. In Ca{\sc ii} it was necessary to use two
Gaussians around $\sim$30~km/s to account for the slightly asymmetric profile of this 
velocity component. Only one Gaussian was necessary to fit the matching velocity 
component of Na{\sc i}. The velocity component at $\sim$40~km/s could
only be detected in the Ca{\sc ii} line. The uncertainty of radial velocities was
estimated by examining the differences between values measured for the same
components in the two lines of the Na{\sc i} and Ca{\sc ii} doublets respectively.

\begin{figure}
\includegraphics[angle=00,width=\hsize]{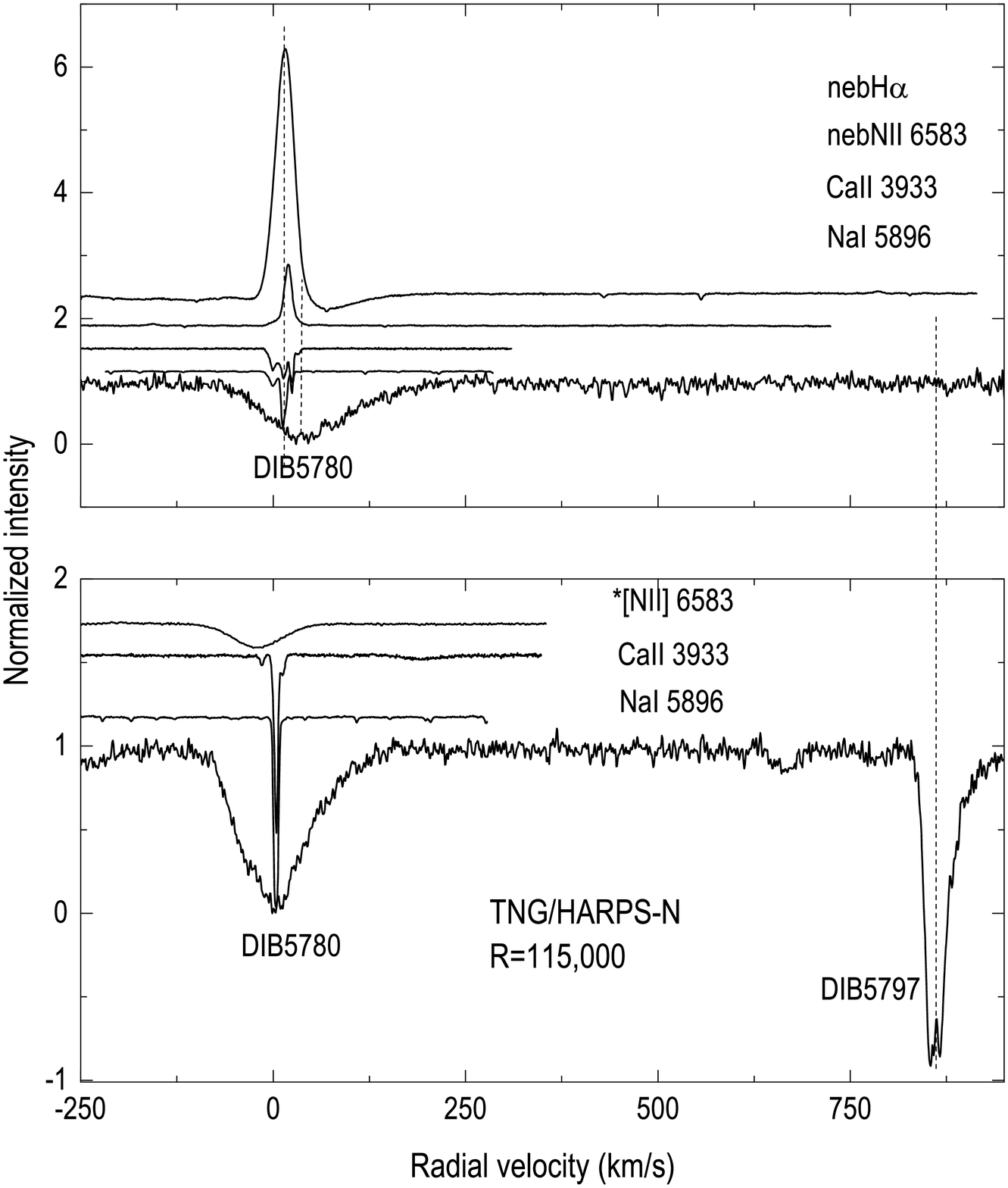}
\caption{Top panel: the strongest DIB at 5780~\AA\ in the
spectrum of HD~37022, shown in radial velocity scale together with other
interstellar and nebular lines. The zero in the velocity scale is chosen to
coincide with the strongest component visible in the D$_1$ Na{\sc i} line.
All spectra have been normalised to the continuum, and shifted for clarity.
Ordinate scale is arbitrary. The two dashed lines mark the position of the
zero velocity component and the position of the peak of the
5780 ~\AA\ DIB, clearly red\textendash shifted by $\sim$30~km/s.
A similar plot is also shown in the bottom panel for the reference
star HD~24398; this object shows no detectable nebular lines. The
5780~\AA\ DIB appears in this case (as in almost all cases) to be
perfectly aligned (within $\sim$2 km/s) with interstellar atomic features.}
\label{dib_shift}
\end{figure}

Figure~\ref{dib_shift} unambiguously shows the red\textendash shift
of the 5780~\AA\ DIB using the HARPS\textendash N spectra. We used the
HD~24398 star as a reference ``normal DIBs'' object. The latter is known as
an object in which all interstellar lines are dominated by a heliocentric
Doppler velocity of about 14~km/s \citep{bondar12}, with some structure
barely detectable only at extremely high resolution and S/N ratio: \citet{welty1996}
resolved 8 components for Ca{\sc ii} with radial velocities between -6 and 22~km/s,
but with $\sim$80\% of the estimated column density being within the range from 13 to 16~km/s.
From this figure it is apparent that the maximum of the
5780~\AA\ DIB in HD~37022 is not found at its expected wavelength, but instead
it is red\textendash shifted by $\sim$0.58~\AA, corresponding to a Doppler
shift of about 30~km/s with respect to the radial velocity of peak absorption
in Na{\sc i}. While the Doppler structure of the Ca{\sc ii} line shows some
red\textendash shifted components, the red shift we measured for the
5780~\AA\ DIB goes clearly beyond it, where no atomic absorption component
was detected. Less evidently, but measurably, the 5780~\AA\ DIB
in HD~37022 is also broader than in HD~24398, by about the same amount.

In the spectrum of HD~37022 we can also detect some nebular emission lines,
likely to arise from the H{\sc ii} region around the same star. The
H$\alpha$ and two forbidden N{\sc ii} lines seen in emission appear to have
a Doppler red\textendash shift of only a few km/s with respect to the
strongest component of the Na{\sc i} lines, in any case well within the
velocity range spanned by the Doppler structure of the Ca{\sc ii}
lines. The H$\alpha$ appears to have a small Doppler shift, a few km/s,
with respect to the N{\sc ii} lines. This agrees with previous studies
of the nebular emission from the Orion Trapezium \citep{baldwin2000}.
Indeed, \citet{baldwin2000} also found much larger velocity differences
among nebular lines due to different species in different ionisation states,
suggesting that they arise from different parts of an accelerating wind.
Since our observations are centered on the trapezium stars, fainter nebular
lines are barely detectable in our spectra, and thus we cannot confirm this.

The H$\alpha$ line also shows a superimposed, much broader absorption
line. Its width hinders an accurate measurement of its radial velocity.
A similar structure, namely an emission line with an underlying, much broader
one in absorption, can be seen in H$\beta$. In H$\beta$ the absorption
component is stronger, and it can be seen to be approximately aligned with
the radial velocity of the emission component.
Upon checking the radial velocity of a few
photospheric lines on our HARPS\textendash N data, (e.~g. N{\sc iii} at
4003.6~\AA,  He{\sc i} at 4026.2~\AA,  C{\sc iii} at 4162.9~\AA, Si{\sc iv}
at 4212.4~\AA, O{\sc ii} at 4366.9~\AA, O{\sc iii} at 5508.1~\AA, and
C{\sc iv} at 5812.0~\AA), they appear consistent with the most current
value found in the literature \citep[$\sim$23.6~km/s][]{olivares2013},
even if there is some scatter between different lines, at the level of less
than $\sim$2~km/s. This scatter may be due to the documented variability
of stellar lines in HD~37022, attributed to the complex interplay of its
magnetic field, fast wind and some infalling material
\citep{simondiaz2006,wade2006}. This is mostly apparent when comparing
e.g. the N{\sc iii} line at 4003.578~\AA\ and the He{\sc i} one at
4009.258~\AA\: the Doppler shift of the former corresponds to a
heliocentric radial velocity of about 9.5 km/s while the latter
of about 41 km/s. However, the vast majority of the observed stellar
lines is consistent with a radial velocity of 22\textendash 23 km/s .

\begin{figure*}
\includegraphics[angle=00, width=14cm]{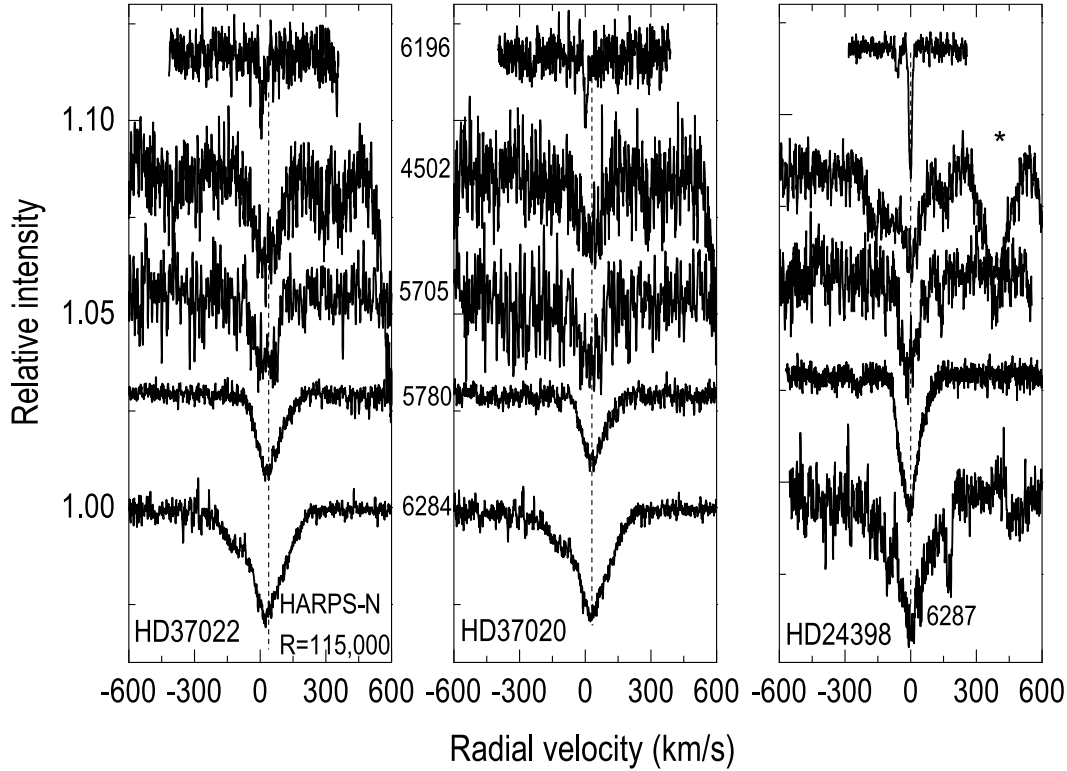}
\caption{Five diffuse bands observed in the HARPS-N spectrum of
$\theta$Ori (left and middle panel) and, for reference, of $\zeta$Per
(right panel). To ease the comparison of their radial velocities,
all DIBs were scaled to the same maximum intensity. All bands have been
referred to the peak absorption in Na{\sc i} lines (see Fig.~\ref{dib_shift}).
Four of them (the relatively broad ones) show a very
similar red\textendash shift in $\theta$Ori. The fifth one \textemdash\
at 6196~\AA\ \textemdash\ is either not red\textendash shifted at all
or the shift is below the level of detection. }
\label{radveldib}
\end{figure*}

We did not include HD~37020 in Fig.~\ref{dib_shift} because it is
almost undistinguishable from that of HD~37022, just with very slightly
weaker DIBs and somewhat lower S/N ratio, hence it would not add any additional
information here (see Fig.~\ref{radveldib}).

Figure~\ref{radveldib} shows a comparison of the DIBs which can be measured in
HD~37020, HD~37022, and the reference star HD~24398 with HARPS\textendash N,
again in radial velocity scale, using the Doppler shift of the strongest
component of the D$_1$ Na{\sc i} line in the same spectrum as reference.
Four of the five DIBs shown, the relatively broad ones, are clearly
red\textendash shifted in HD~37020 and HD~37022 with respect to the
D$_1$ Na{\sc i} line marking the zero of the velocity scale. Their shift is
compatible with a unique relative radial velocity, and all of them also
exhibit a similar broadening with respect to the DIBs measured in the
reference star HD~24398. Rather remarkably,
the peak of the narrow 6196~\AA\ DIB in HD~37020 and HD~37022 appears to be
perfectly aligned (within $\sim$5 km/s) with the Doppler velocity of 
the atomic lines. The ratio
of DIB intensity is also evidently different between HD~37022 and HD~24398:
in the former the 6284~\AA\ DIB is stronger than the 5780~\AA\ one, while
the reverse happens in the latter. Morever, a weak, narrow DIB at
6287~\AA\ appears on the red wing of the 6284~\AA\ DIB in
HD~24398, that is below the level of detection in HD~37022.


\begin{figure}
\includegraphics[angle=00,width=\hsize]{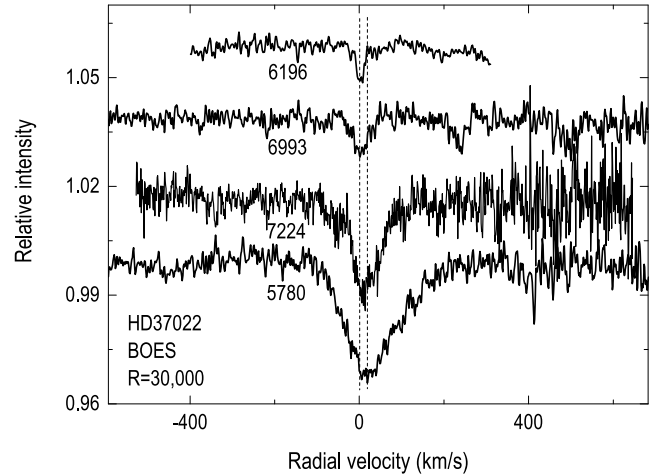}
\caption{Four DIBs from the BOES spectra of HD~37022, including two
not measurable with HARPS\textendash N. The narrow 6993~\AA\ feature
is evidently unshifted as the 6196\AA\ DIB, while the broad 7224~\AA\ one
shares the same red\textendash shift of the 5780~\AA\ DIB. } \label{irdib}
\end{figure}

To further strengthen that the observed redshift is not an (unlikely)
instrumental artifact of HARPS\textendash N, Fig.~\ref{irdib} shows the
5780 and 6196~\AA\ DIBs observed in HD~37022 with BOES at slightly lower
spectral resolution (90,000 vs 115,000), which precisely match  
(within $\sim$1 km/s) the HARPS\textendash N data. In addition, the same 
Fig.~\ref{irdib} also shows two more DIBs that are out of the spectral range 
covered by HARPS\textendash N, namely the 6993 and 7224~\AA\ DIBs. The peak 
of the former, relatively narrow, precisely matches (within $\sim$1 
km/s) the Doppler shift of the D$_1$ Na{\sc i} line, like the 6196~\AA\ DIB. 
The 7224~\AA\ DIB, conversely, a relatively broad one, appears again 
red\textendash shifted like the 5780~\AA\ DIB.

For a solid quantitative measurement of the radial velocity shifts 
visually apparent in Figs.~\ref{dib_shift}, \ref{radveldib}, and \ref{irdib},
we used the same method commonly used to precisely measure the radial 
velocities of stars, correlating their observed spectra with a template 
\citep[see e.~g.][ and references therein]{Queloz1995}. 
We performed the correlation analysis one DIB at a time, using as
a template the DIB profiles measured in HD~24398. This measurement is not 
coincident with velocity shifts measured using DIB peaks as a reference, nor is 
it expected to be: the correlation yields velocity shifts between feature centroids, 
which would coincide with shifts measured on the peaks only in case of infinite 
signal to noise ratio and perfectly identical spectral shapes between spectrum and 
correlation template. Most DIBs are detectably broader in the Orion Trapezium with 
respect to HD~24398, and noise is clearly far from negligible. Rather remarkably, 
the only two narrow DIBs, for which the radial velocity shift thus determined is 
most significant and least sensitive to the determination method, are compatible 
with zero radial velocity relative to the ``interstellar rest frame'' given by the 
strongest components in the Na{\sc i} D$_1$ and CH 4300 \AA\ lines.
This correlation procedure, as implemented both in the DECH and IRAF data reduction 
softwares, yields the radial velocities maximising the correlation and their
standard deviations, via R\textendash statistics analysis.
The results are listed in Table~\ref{rv_shifts}.

\begin{table}
\caption{Radial velocity displacement (in km/s) of the DIBs, obtained by
correlating their spectra in the Orion Trapezium stars with the reference
spectra observed towards HD~24398. Radial velocities for the Trapezium 
stars are referred to the strongest D$_1$ line component, assumed to mark the 
``interstellar rest frame''. For HD~24398 the ``interstellar rest frame'' is
assumed to coincide with the radial velocity of the molecular CH 4300 \AA\ line.
The listed rest wavelengths for the diffuse bands are from \citep{GMKW00}.
}
\label{rv_shifts}
\begin{center}
\begin{tabular}{lrr}
\hline 
\hline 
 DIB          & HD~37022    & HD~37020    \\
\hline
4501.80~\AA       & +25$\pm$10~km/s  & +30$\pm$10~km/s  \\
5705.20~\AA       & +28$\pm$5~km/s   & +33$\pm$5~km/s   \\
5780.37~\AA       & +24$\pm$1~km/s   & +23$\pm$1~km/s   \\
6195.97~\AA       &  +3$\pm$4~km/s   &  +6$\pm$5~km/s   \\
6283.85~\AA       & +20$\pm$1~km/s   & +22$\pm$1~km/s   \\
6993.18~\AA       &  -1$\pm$5~km/s   &     n/a     \\
7223.91~\AA       & +27$\pm$1~km/s   &     n/a     \\
\hline 
\hline
\end{tabular}
\end{center}
\end{table}

\section{Discussion and conclusions}

The red\textendash shift of the DIBs at 4502, 5705, 5780, 6284, and
7224~\AA\ observed in HD~37022 can be interpreted in two different ways:
either the peculiar physical conditions along, or around, HD~37022 produce
\emph{intrinsically} different DIB profiles, broader and peaking at a
different position, or the observed DIBs are produced by Doppler structure,
as possibly hinted by all shifted DIBs sharing (or compatible with) the
same red\textendash shift and broadening in velocity scale.

\begin{figure}
\includegraphics[angle=00,width=\hsize]{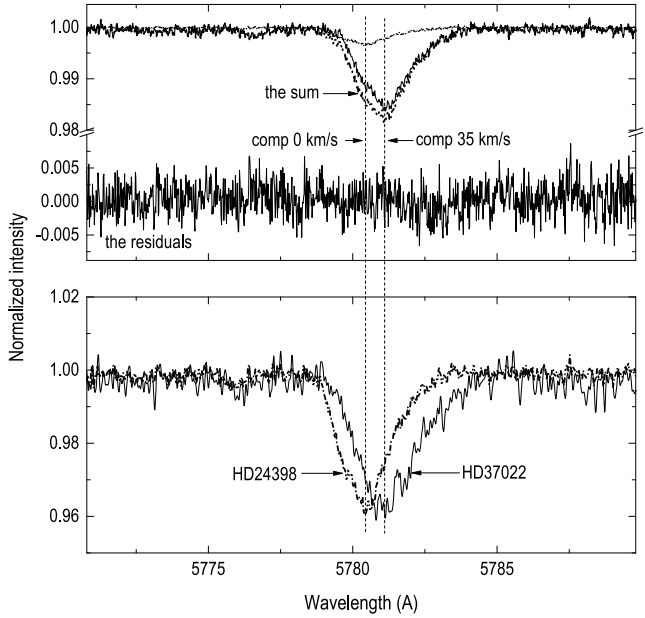}
\caption{Upper panel: the 5780~\AA\ DIB profile in HD~37022, fitted with the
composition of two profiles identical to that of HD~24398. The radial
velocity difference is 35~km/s. Lower panel: profile difference between the
``standard'' reference one of HD~24398 and the one observed in HD~24398, both
normalized to the same peak depth and referred to the reference frame given by
the radial velocity of the peak absorption in the Na{\sc I} atomic lines
observed in their respective spectra.} \label{fit5780}
\end{figure}

\begin{figure}
\includegraphics[angle=00,width=\hsize]{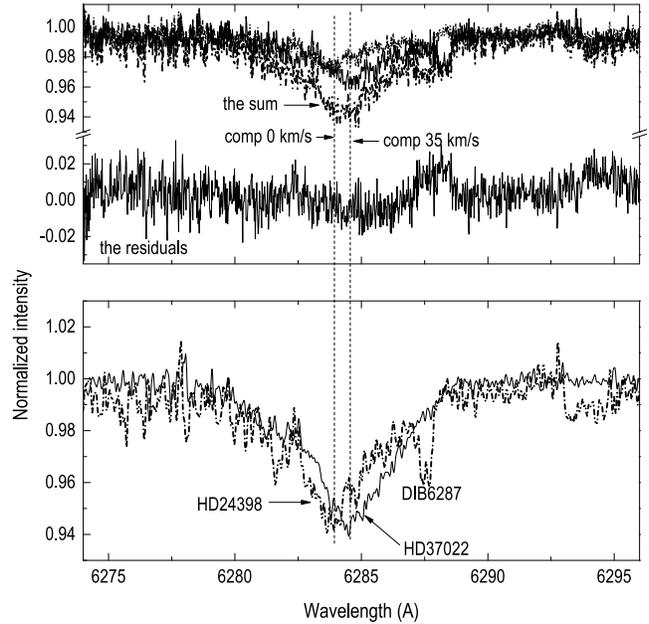}
\caption{Same as in Fig~\ref{fit5780}, for the 6284~\AA\ DIB. The radial
velocity difference between the two components is the same, namely 35~km/s.
Note the presence, only in HD~24398, of the narrow 6287 DIB.} \label{fit6284}
\end{figure}

To test the feasibility of this latter hypothesis, we attempted a conditioned
least\textendash squares fit of the red\textendash shifted 5780 and
6284~\AA\ DIBs in HD~37022 with a superposition of two ``template'' DIBs,
namely the ones observed in HD~24398. Therefore, the DIB as observed in
HD~24398 was assumed as the ``intrinsic'' profile, and several
copies of it with different scalings and Doppler shifts were used to fit
the profile observed in HD~37022. The fit was constrained to only two Doppler
components, one fixed at the Doppler velocity of the Na{\sc i} lines,
the other free to vary. The free parameters of the fit are therefore only
three (for each DIB): the two intensities of the components (constrained to
be positive) and the Doppler velocity of the red\textendash shifted component.
We used the well\textendash known Levemberg\textendash Marquardt method
\citep{levenberg1944}, as implemented in the MPFIT
package \citep{markwardt2009}\footnote{http://purl.com/net/mpfit} for
IDL. This algorithm strives to numerically reduce the $\chi^2$ computed
between model and data, iteratively modifying an initial guess for the
values of the free parameters, until a minimum is found. It is therefore
possible to fall into local minima, especially with noisy data, and the
solution found will in this case depend on the initial starting point.
When several minima exist, they will normally have different $\chi^2$
values, and some may be rejected on the sole basis of this. However, it
may happen that more than one solution has an acceptable $\chi^2$ value;
in this case, some of these may be rejected because the residual
(difference between model and data) is clearly not white noise, i.~e.
with the model falling systematically below the data in some section of
the spectrum and above it in another.
This indeed happens with our fit here, with several solutions being in
principle acceptable on the basis of $\chi^2$ alone, but with only the best
one (i.e. the one with the absolute minimum $\chi^2$ value) producing a
residual of white noise, with uncorrelated adjacent values.
To make the determination of the acceptable fit more robust and less
subjective, we filtered both the HD~37022 and the template DIBs with a
mild smoothing filter (Fourier 5 points filter as implemented in the
DECH data reduction suite). This is justified by the fact that
HARPS\textendash N spectra are oversampled and the DIBs being fitted are
intrinsically rather broad. When using the filtered data,
only one acceptable solution is found, which corresponds to the
absolute minimum of $\chi^2$, while the unphysical ones are not even local
minima any more. The resulting, very satisfactory best fits for the
5780 and the 6284~\AA\ DIBs are shown in Figs.~\ref{fit5780} and \ref{fit6284}.

\begin{figure}
\includegraphics[angle=00,width=\hsize]{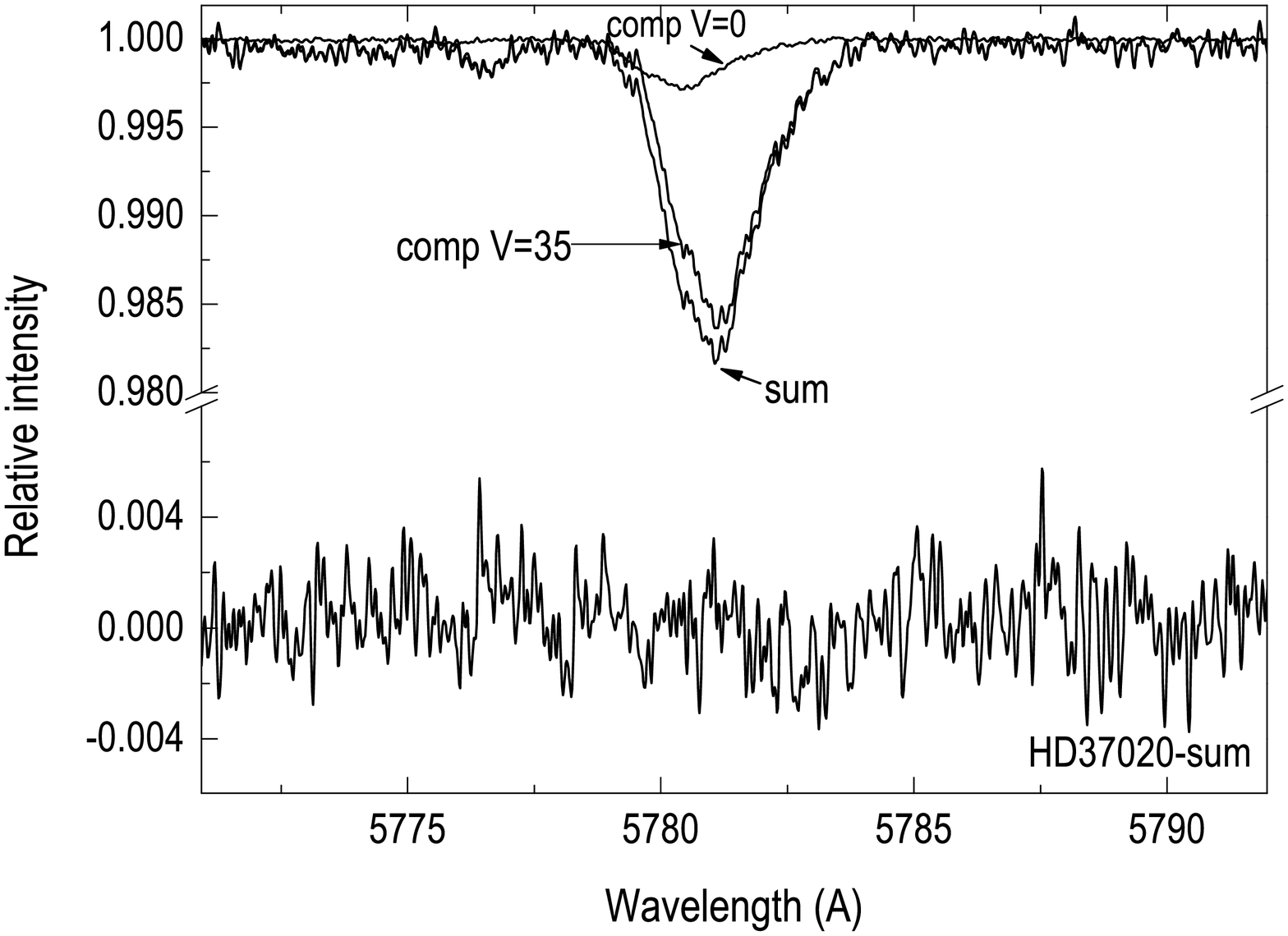}
\caption{The 5780~\AA\ DIB profile in HD~37020, fitted with the composition
of two profiles identical to that of HD~24398 (same as the upper panel in
Fig.~\ref{fit5780}. The radial velocity difference is the same, namely 35~km/s.}
\label{fit5780_37020}
\end{figure}

\begin{figure}
\includegraphics[angle=00,width=\hsize]{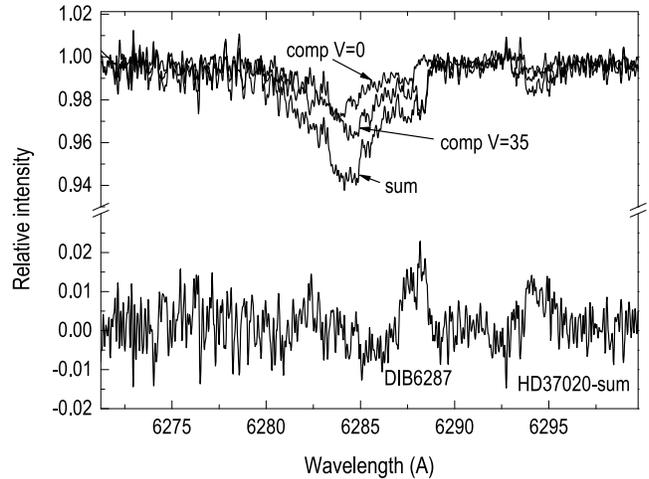}
\caption{The 6284~\AA\ DIB profile in HD~37020, fitted with the composition
of two profiles identical to that of HD~24398 (same as the upper panel in
Fig.~\ref{fit6284}. The radial velocity difference is the same, namely 35~km/s.}
\label{fit6284_37020}
\end{figure}

In both cases, the residual between the fit and the original spectrum
is essentially white noise, except for the obvious absence of the
narrow DIB at 6287~\AA\ in HD~37022, which is instead present in the
reference HD~24398 spectrum and thus appears as an artifact in the
residual in Fig.~\ref{fit6284}. When slightly smoothed spectra are
used for the fit, the solution is very robust for both the 5780 and
6284~\AA\ DIBs, and in both cases the additional fitted component
results to be red\textendash shifted with respect to interstellar sodium
by 34~km/s~$\leq v_\mathrm{add} \leq$~40~km/s, the best solution being
$\sim$35~km/s. When the same fitting procedure is applied to the
5780 and 6284~\AA\ DIBs in HD~37020, due to lower S/N the solution is
not unique even after smoothing, but all acceptable solutions yield
nonetheless 34~km/s~$\leq v_\mathrm{add} \leq$~40~km/s for the
radial velocity of the additional component, consistent with what
was found for HD~37022. These results can be compared to the
radial velocity shifts listed in Table~\ref{rv_shifts}. However, 
since the latter were determined by correlation with a template 
consisting of a single component, for a meaningful comparison one 
must compare them with the weighted average of the radial velocities 
of the two components of the fit performed here, the weights being their
relative intensities. This results in very nearly the same numbers 
of Table~\ref{rv_shifts} for the DIBs at 5780 and 6284~\AA.
The noise in the other DIBs in HD~37022 and HD~37020 is so high that the
fits are not robust (i.~e. many equivalently acceptable fit solutions
are found) but they are compatible with the results found for the 5780
and 6284~\AA\ DIBs (i.~e. $v_\mathrm{add} \sim 35$~km/s is always one of
the acceptable solutions of the fit). Therefore,
while of course we do not claim that this is the only possible
interpretation of the data, we do find it to be possible and
self\textendash consistent.

Of course, this then raises more questions:
what is producing the redshifted component of the DIBs?
Some residual, still infalling material of the young,
newly\textendash formed, very early\textendash type stars?
Or do they arise in the evaporating layer of the PDR, at the edge of
the H{\sc ii} region? And why is this additional Doppler component
not detected in atomic lines? Upon comparing the Doppler
structures of the Na{\sc i} and Ca{\sc ii} lines in
Fig.~\ref{atom_doppler} and Table~\ref{atomic_radvel_table},
there appears to be a temperature gradient with increasing radial
velocities, whereby Na{\sc i} lines decrease while Ca{\sc ii} increases,
and the highest velocity component is detected only in Ca{\sc ii},
but weaker. There may therefore be another, more red\textendash shifted
component with a still higher temperature, in which thus
sodium and singly ionized calcium are absent, but
this is a long stretch with the available data. Observations
with a substantially higher S/N ratio would enable us to perform a
robust fit on the other three red\textendash shifted DIBs: it is clear
that if 5 different DIBs were \emph{robustly} consistent with
\emph{the same} Doppler structure, yielding the same radial velocity
as the only acceptable fit for the additional component of all of
them, this would provide rather solid evidence for this
interpretation. Conversely, if each DIB were found to
require a different radial velocity structure, this would make this
interpretation just a pointless mathematical exercise.
Furthermore, a substantially higher S/N ratio might also enable us to
detect either additional Doppler components in sodium and calcium lines,
or weak molecular features (CH, CH$^+$, CN), which are below the detection
limit in current data.
This would be extremely interesting also because there is some
observational evidence that e.~g.  CH, CH$^+$ and CN features do not
always show all the same Doppler components, with CH$^+$ features
sometimes shifted with respect to the other species and to atomic
lines \citep[see e.~g.][]{allen1994}.

As to the other possible interpretation, i.~e. different intrinsic
profiles of the red\textendash shifted DIBs in HD~37022, this is tempting
in the light of the recent observations in Herschel~36
\citep{york2014,oka2014,oka2014b}.
In this extremely peculiar line of sight, CH and CH$^+$
were observed to have an anomalously high rotational temperature, and
some DIBs were simultaneously found to be anomalously broad (much more
than what we observe here), with a very wide red wing, and the peaks of
some of them slightly redshifted (less than what we observe here). This
has been tentatively interpreted as unresolved rotational envelopes of
molecular bands, which in Herschel~36 appear to have an anomalously
high rotational excitation.
If this were the correct interpretation for HD~37022, then this would be
an intermediate case between the ``standard'' excitation conditions of
the cold neutral ISM, where ``normal'' DIBs arise, and those prevailing
in Herschel 36. However, the DIB at 5797~\AA\ is prominent, and
prominently broadened, in Herschel~36, and instead is undetectable in
HD~37022; also, the 6196~\AA\ DIB is one of the broadened ones in
Herschel~36, while no such effect is evident in our data of HD~37022 (but
\emph{some limited} broadening might be consistent with our observations,
hidden by the noise). Also, the 6284~\AA\ DIB is broadened but not redshifted
in Herschel~36, while it \emph{is} redshifted in HD~37022.
To test this hypothesis, again it would be useful to obtain deeper
observations with higher S/N, to detect CH, CH$^+$ and CN features,
determine their rotational temperatures in HD~37022 and explore their relations
with the observed profiles and positions of red\textendash shifted DIBs.
We remark that HD~37022 and Herschel~36 also have rather peculiar extinction
curves \citep{FM07}, hinting that dust properties are
also unusual. Actually, Herschel~36 appears to be a tight triple system
\citep{arias2010,sanchez2014}. \citet{FM07} give for it the spectral
classification of the third, most luminous, O~7.5~V type star of the system,
which certainly dominates the IUE spectrum they used to derive the extinction.
The spectrum in the visible, in which the peculiar DIBs were observed,
contains non\textendash negligible contributions from all three unresolved
stars.
In any case, the extinction curve and DIBs are very likely to be extremely
similar for all three stars of such a close system, unless circumstellar
material makes a large contribution to them.

\begin{figure}
\begin{center}
\includegraphics[angle=00,width=6.9cm]{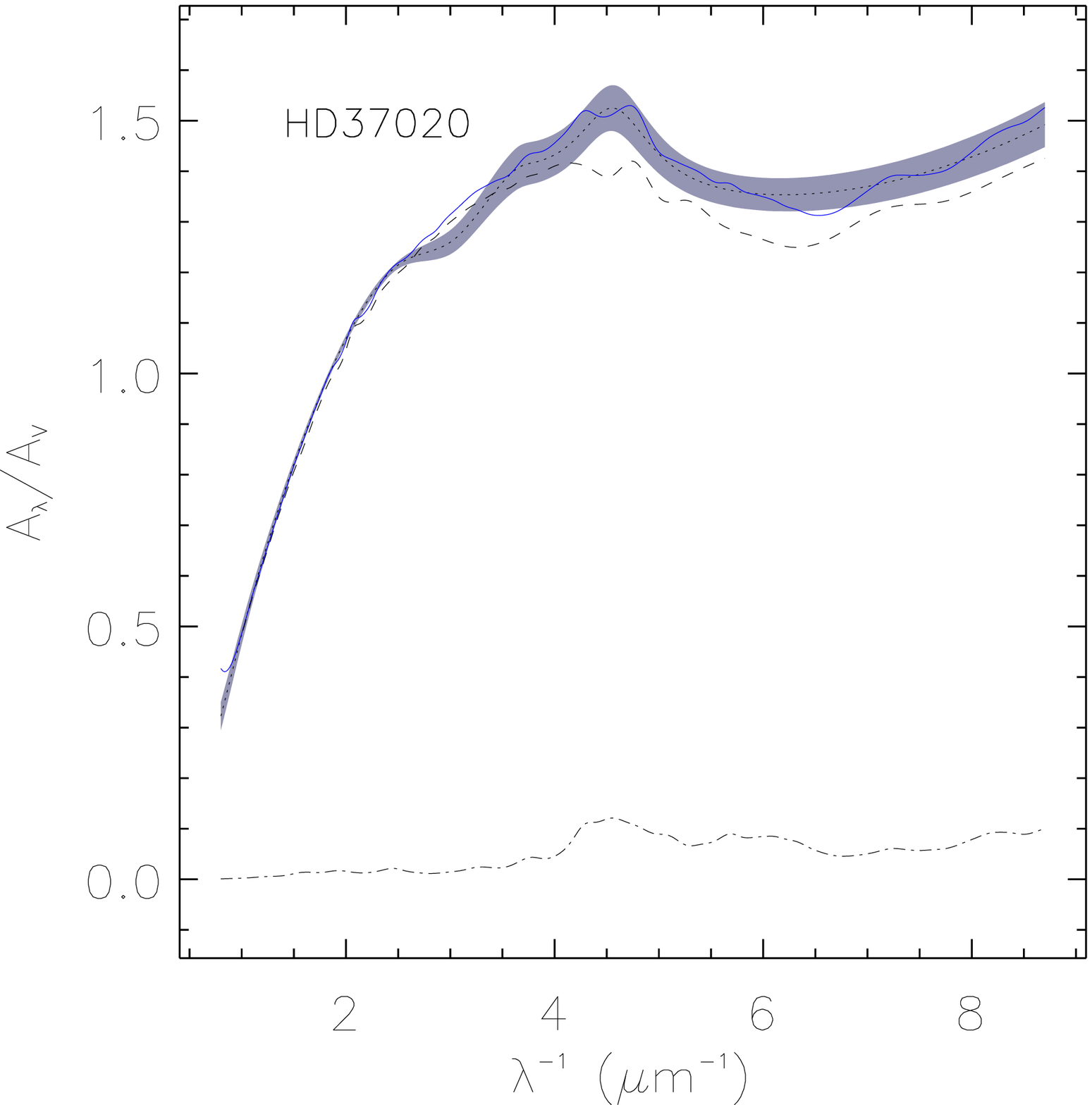}
\includegraphics[angle=00,width=6.9cm]{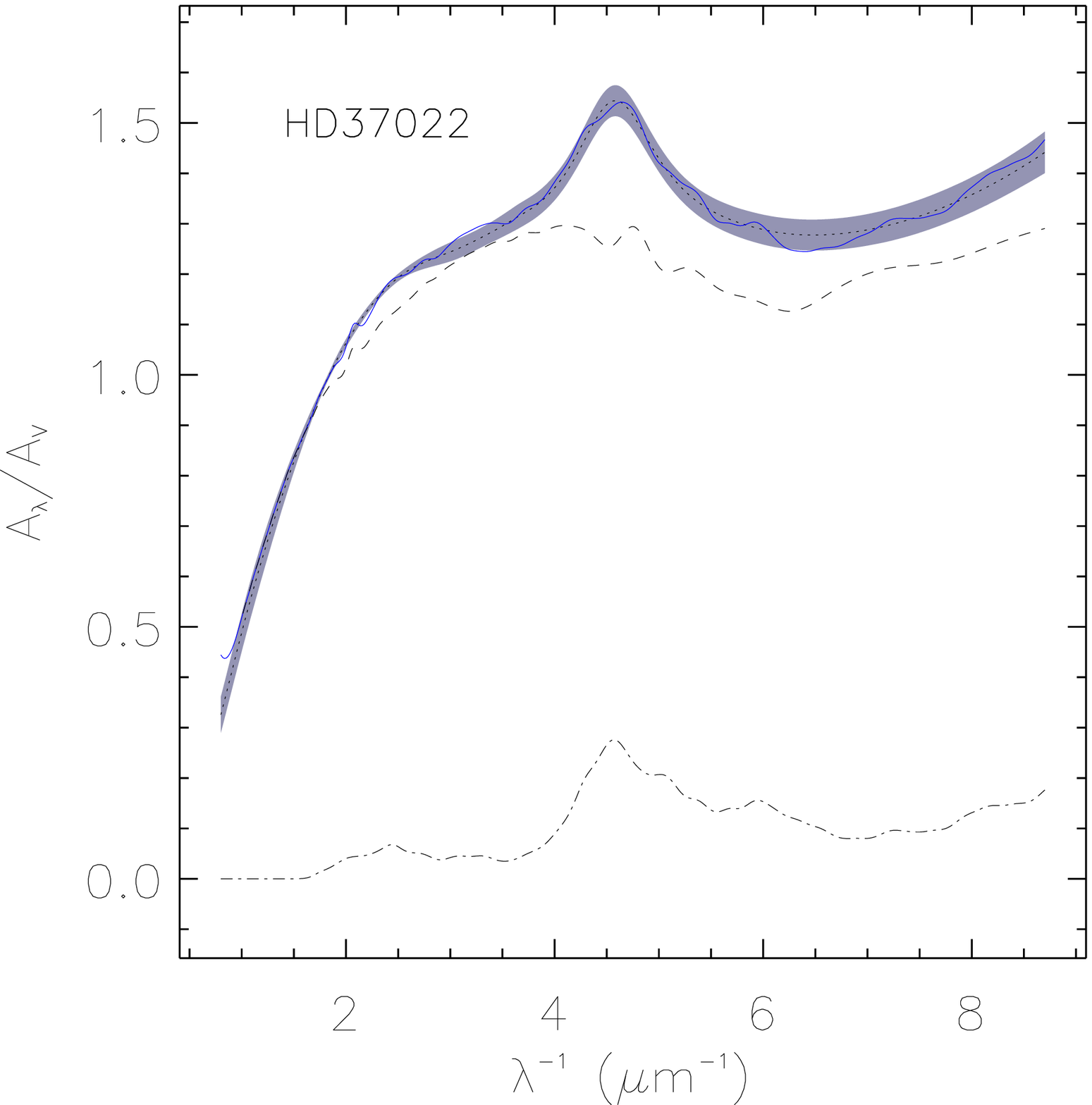}
\includegraphics[angle=00,width=6.9cm]{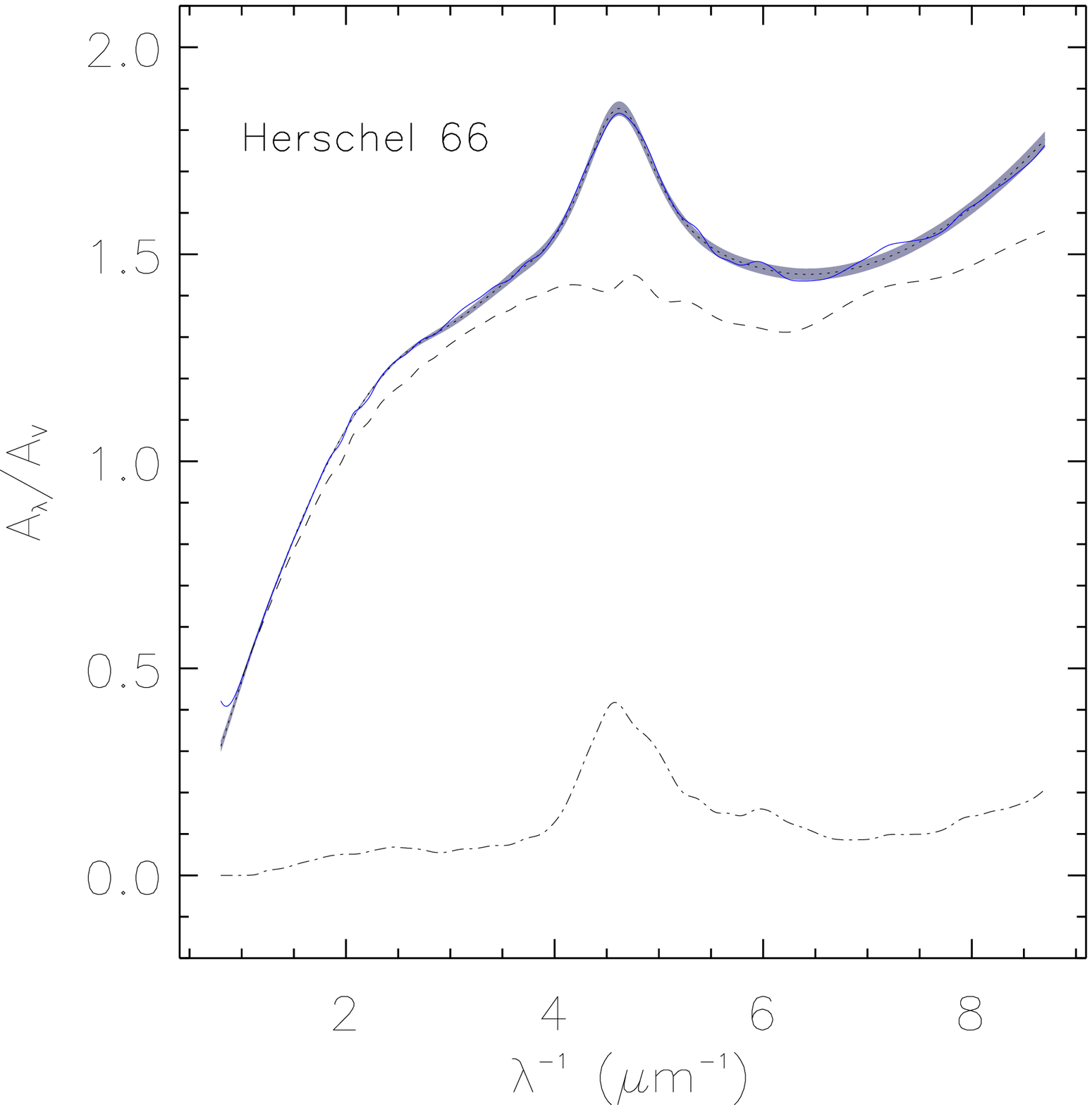}
\end{center}
\caption{Extinction curves of HD~37020, HD~37022, and Herschel~36.
The dotted lines are the observations by \citet{FM07}, the shaded area represents
the error estimated by the authors. The continuous lines represent the best fit
with the [CM]$^2$ model \citep{mulas2013}, the dashed lines the contribution by
macroscopic dust, the dash\textendash dotted lines the contribution by PAHs
\citep{mulas2013}.
} \label{extinctioncurves}
\end{figure}

\begin{figure}
\begin{center}
\includegraphics[angle=00,width=6.9cm]{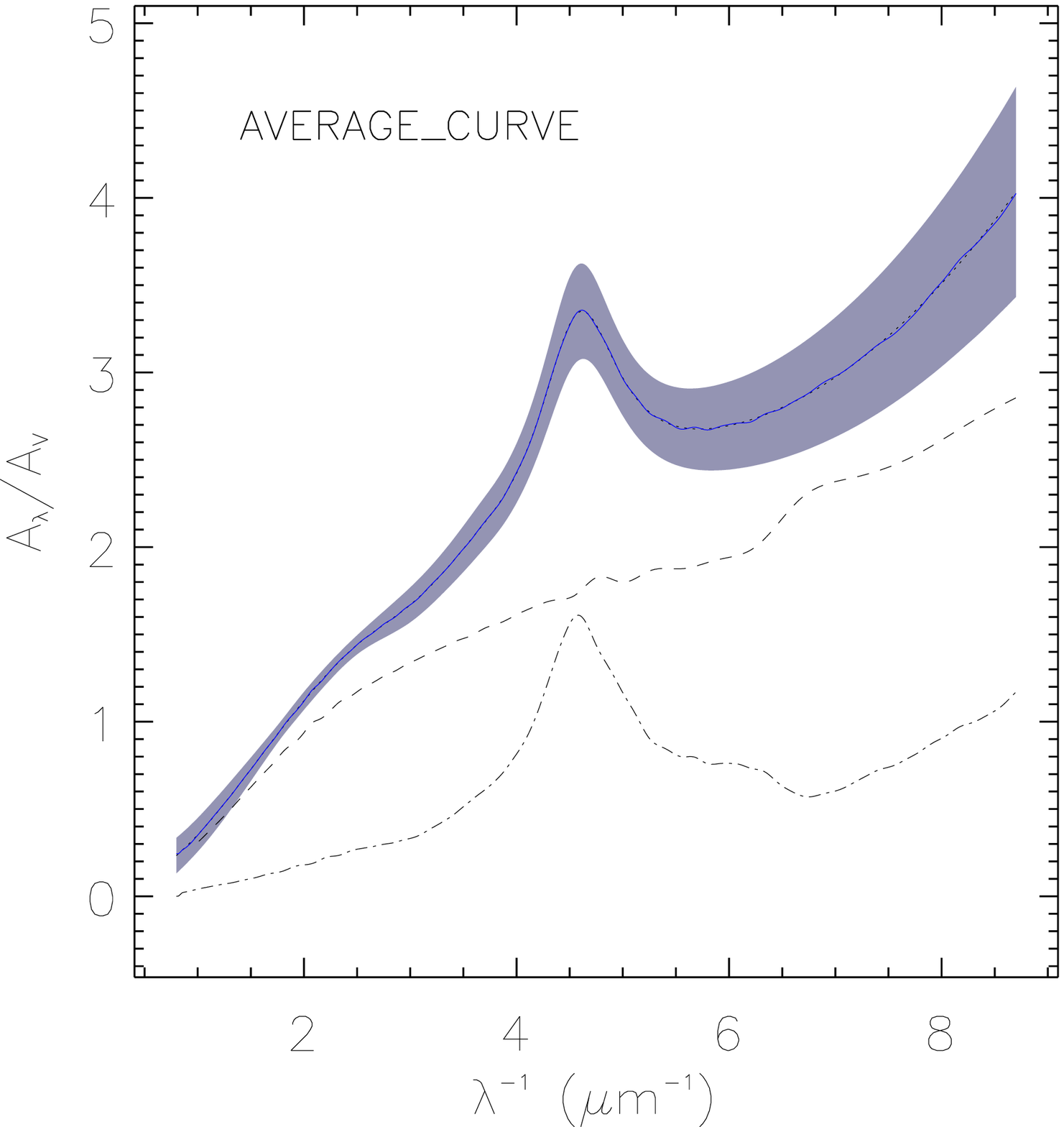}
\includegraphics[angle=00,width=6.9cm]{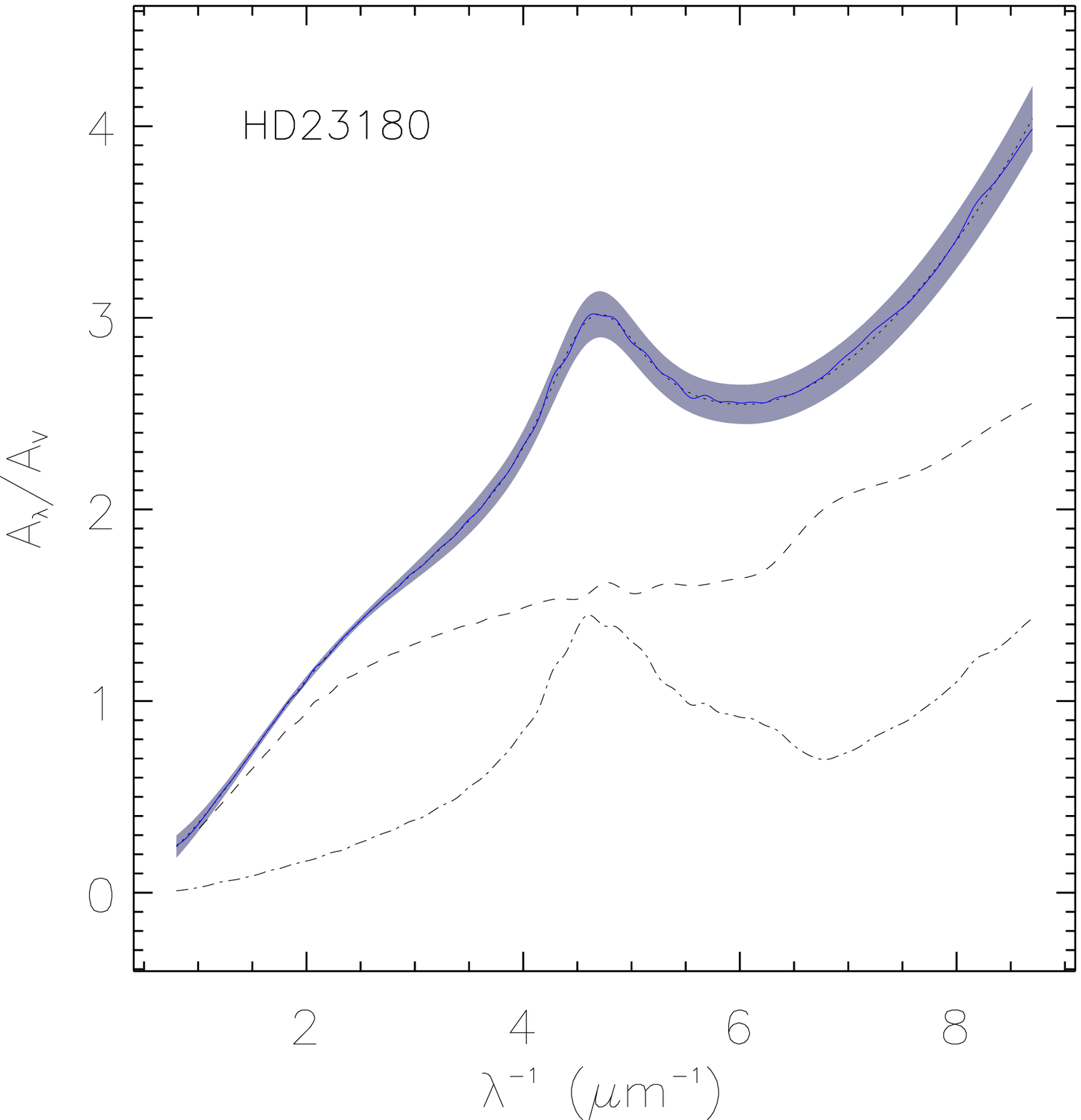}
\end{center}
\caption{Same as Fig.~\ref{extinctioncurves}, but for the average Galactic
extinction curve and HD~23180 ($o$~Per), another star in the same association
of HD~24398 ($\zeta$~Per), likely to have a very similar extinction curve.
} \label{extinctioncurvesref}
\end{figure}
Figure~\ref{extinctioncurves} shows the extinction curves
reported by \citet{FM07} for HD~37020, HD~37022, and Herschel~36,
together with their fits with the [CM]$^2$ dust model \citep{mulas2013}.
Figure~\ref{extinctioncurvesref}, for comparison, shows the extinction curve
of HD~23180 ($o$~Per), the closest proxy of HD~24398 ($\zeta$~Per) available
in the sample of \citet{FM07}, and the average galactic one.
The macroscopic dust component appears to be very similar for all the three
stars in Fig.~\ref{extinctioncurves}, all very different from the
reference ones in Fig.~\ref{extinctioncurvesref}. Upon examining Tables~4
and 5 in \citet{mulas2013}, one sees that in all three cases, upon comparison
with the galactic average and HD~23180, the dust population is dominated by
large silicatic grains, covered by a thick mantle of heavily processed
(entirely, or almost entirely, aromatic) carbonaceous material.
PAHs are very underabundant in all of them with respect to the reference ones,
similar to the galactic average. In the evolutionary scenario
put forward by \citet{cecchi2014}, all three lines of sight appear to be
composed of ``old'', heavily processed dust.

It is unclear whether both DIBs and the extinction curves are made
peculiar by some common physical cause related to the environment, or if
DIBs may be altered due to particular dust properties. Red shift and
broadening in molecular electronic bands are commonly caused by attachment
of molecules to a solid substrate \citep[see e.~g.][]{tielens1987}.
In an environment apparently devoid of small dust particles and PAHs
\citep{mulas2013}, in which apparently smaller nanograins and
macromolecules are either destroyed or coalesce onto larger ones,
possibly some DIB carriers can also partially attach to grain surfaces,
producing what we observe in HD~37022.

Clearly a strategy to pursue to try to find the cause of the red shift
of some DIBs in HD~37022, besides obtaining better observations of
HD~37022 and valid regardless of the interpretation, would be to find
other similar cases, in order to understand what they have in common and
thus what makes them different from the ``standard'' ones. Selecting
lines of sight for further DIB observations choosing the ones with
peculiar extinction curves similar to that of HD~37022
\citep{FM07} might be a key in this respect.

\section*{Acknowledgments}
This paper includes data gathered with the 3.5-m Telescopio Nazionale
Galileo, and data collected with the 1.8 m telescope at Bohyunsan Optical
Astronomy Observatory (South Korea). \\
JK acknowledges the financial support of the Polish National
Center for Science during the period 2012 - 2015 (grant
UMO-2011/01/BST2/05399). \\ GAG acknowledges the support of Chilean
fund FONDECYT-regular (project 1120190). \\ GM and CCP acknowledge the
support of the Autonomous Region of Sardinia, Project CRP 26666 (Regional
Law 7/2007, Call 2010)

\label{lastpage}
\end{document}